\documentclass[12pt,a4paper]{article}
\makeatletter
\def\eqnarray{%
\stepcounter{equation}%
\let\@currentlabel=\theequation
\global\@eqnswtrue
\global\@eqcnt\z@
\tabskip\@centering
\let\\=\@eqncr
$$\halign to \displaywidth\bgroup\@eqnsel\hskip\@centering
$\displaystyle\tabskip\z@{##}$&\global\@eqcnt\@ne
\hfil$\displaystyle{{}##{}}$\hfil
&\global\@eqcnt\tw@$\displaystyle\tabskip\z@{##}$\hfil
\tabskip\@centering&\llap{##}\tabskip\z@\cr}
\makeatother
\newcommand{\ket}[1]{{\vert{#1}\rangle}}
\newcommand{\bra}[1]{{\langle{#1}\vert}}
\newcommand{\kett}[1]{{\vert{#1}\rangle\rangle}}
\newcommand{\braa}[1]{{\langle\langle{#1}\vert}}
\newcommand{\calh}{{\cal H}}

\newcommand{\fukuso}{{\mathbf C}}
\newcommand{\futon}{{\bf N}}

\newcommand{\zetta}{{\vert z\vert}}

\newcommand{\kappazetta}{{\vert\kappa\vert}}

\begin{document}

\title{\sl N--Level System Interacting with Single Radiation Mode 
           and Multi Cat States of Schr{\" o}dinger in the Strong 
           Coupling Regime}
\author{
  Kazuyuki FUJII
  \thanks{E-mail address : fujii@yokohama-cu.ac.jp }\\
  Department of Mathematical Sciences\\
  Yokohama City University\\
  Yokohama, 236-0027\\
  Japan
  }
\date{}
\maketitle\thispagestyle{empty}
%
%
%
%
\begin{abstract}
  In this paper we consider some (generalized) models 
  which deal with n--level atom interacting with a single radiation 
  mode and study a general structure of Rabi floppings of these 
  generalized ones in the strong coupling regime. 
  
  To solve these models we introduce multi cat states of 
  Schr{\" o}dinger in our terminology 
  and extend the result in (quant--ph/0203135) which Rabi frequencies 
  are given by matrix elements of generalized coherent operators 
  (quant--ph/0202081) under the rotating--wave approximation. 
  Our work is a full generalization of Frasca (quant--ph/0111134) and 
  Fujii (quant--ph/0203135). 
  
  In last we make a brief comment on an application to Quantum Computation 
  on the space of qudits. 
\end{abstract}

\newpage

%
%
%
%

\section{Introduction}

Coherent states or generalized coherent states play an important role in 
quantum physics, in particular, quantum optics, see \cite{KS} and \cite{MW}. 
They also play an important one in mathematical physics. See 
the textbook \cite{AP}. For example, they are very useful in performing 
stationary phase approximations to path integral, \cite{FKSF1}, 
\cite{FKSF2}, \cite{FKS}. 

Coherent operators which produce coherent states are very useful because 
they are unitary and easy to handle. 
The basic reason is probably that they are subject to the elementary 
Baker-Campbell-Hausdorff (BCH) formula. Many basic properties of them 
are well--known, see \cite{AP} or \cite{KF6}. 

Generalized coherent operators which produce generalized coherent states 
are also useful. But they are not so easy to handle in spite of having  
the disentangling one corresponding to the elementary BCH formula. 
In \cite{KF5} and \cite{KF14} the author determined all matrix elements 
of generalized coherent operators based on Lie algebras su(1,1) and su(2). 
They are interesting by themselves, but moreover have a very interesting 
application. 

In \cite{MFr} Frasca dealt with a model that  
describes a 2--level atom interacting with a single radiation mode 
(see \cite{MSIII} for a general review) in the strong coupling regime 
(not weak coupling one !) and showed that Rabi frequencies are obtained by 
matrix elements of coherent operator under the rotating--wave approximation. 
His aim was to explain the recent experimental finding on Josephson 
junctions \cite{NPT}. 

This is an interesting result and moreover his method can be widely 
generalized. See also \cite{MFr2}, \cite{MFr4} for another examples 
dealt with in the strong coupling regime. 

His model has been generalized by Fujii \cite{KF15} 
by making use of some operators based on Lie algebras su(1,1) and su(2), 
and a mathematical structure of Rabi floppings of these extended 
models in the strong coupling regime has been studied. 

These models are based on 2--level atom interacting with a "single" 
radiation mode $\cdots$ 2--level system in the following. 
By the way, we are interested in n--level atom interacting with a "single" 
radiation mode $\cdots$ n--level system $\cdots$ which has been inspired by 
Quantum Computation, \cite{MSJr}, \cite{CBKG}, \cite{KBB}, \cite{NBD}, 
\cite{KF16}, \cite{KuF}.  
That is, we would like to extend the results in 2--level system to ones in 
n--level system. This extension is not so easy. 

In this paper we construct (propose) models that describe a n--level atom 
interacting with a "single" radiation mode. 
To solve these models we introduce multi cat states of Schr{\" o}dinger 
(in our terminology) and show that (generalized) Rabi frequencies are also 
given by matrix elements of generalized coherent operators under 
the rotating--wave approximation. 

We believe that the results will give a new aspect to Quantum Optics or 
Mathematical Physics. 

Lastly we discuss an application to Quantum Computation on the space of 
qudits and discuss a possibility of application to Holonomic Quantum 
Computation, \cite{ZR}, \cite{PZR}, \cite{PC}, \cite{PZ}, 
\cite{KF0}--\cite{KF4}, but our discussion is not complete. 
See also Discussion in \cite{KF15} concerning this possibility.

\section{Coherent and Generalized Coherent Operators}

\subsection{Coherent Operator}
Let $a(a^\dagger)$ be the annihilation (creation) operator of the harmonic 
oscillator.
If we set $N\equiv a^\dagger a$ (:\ number operator), then
\begin{equation}
  \label{eq:2-1-1}
  [N,a^\dagger]=a^\dagger\ ,\
  [N,a]=-a\ ,\
  [a^\dagger, a]=-\mathbf{1}\ .
\end{equation}
Let $\calh$ be a Fock space generated by $a$ and $a^\dagger$, and
$\{\ket{n}\vert\  n\in\futon\cup\{0\}\}$ be its basis.
The actions of $a$ and $a^\dagger$ on $\calh$ are given by
\begin{equation}
  \label{eq:2-1-2}
  a\ket{n} = \sqrt{n}\ket{n-1}\ ,\
  a^{\dagger}\ket{n} = \sqrt{n+1}\ket{n+1}\ ,
  N\ket{n} = n\ket{n}
\end{equation}
where $\ket{0}$ is a normalized vacuum ($a\ket{0}=0\  {\rm and}\  
\langle{0}\vert{0}\rangle = 1$). From (\ref{eq:2-1-2})
state $\ket{n}$ for $n \geq 1$ are given by
\begin{equation}
  \label{eq:2-1-3}
  \ket{n} = \frac{(a^{\dagger})^{n}}{\sqrt{n!}}\ket{0}\ .
\end{equation}
These states satisfy the orthogonality and completeness conditions
\begin{equation}
  \label{eq:2-1-4}
   \langle{m}\vert{n}\rangle = \delta_{mn}\ ,\quad \sum_{n=0}^{\infty}
   \ket{n}\bra{n} = \mathbf{1}\ . 
\end{equation}

\noindent{\bfseries Definition}\quad We call a state  
\begin{equation}
\label{eq:2-1-7}
\ket{z} =  \mbox{e}^{za^{\dagger}- \bar{z}a}\ket{0}\equiv U(z)\ket{0} 
\quad \mbox{for}\quad z\ \in\ \fukuso 
\end{equation}
the coherent state.

\subsection{Generalized Coherent Operator Based on $su(1,1)$}
Let us state generalized coherent operators and states based on $su(1,1)$.

We consider a spin $K\ (> 0)$ representation of $su(1,1) 
\subset sl(2,\fukuso)$ and set its generators 
$\{ K_{+}, K_{-}, K_{3} \}\ ((K_{+})^{\dagger} = K_{-})$, 
\begin{equation}
  \label{eq:2-2-3}
 [K_{3}, K_{+}]=K_{+}, \quad [K_{3}, K_{-}]=-K_{-}, 
 \quad [K_{+}, K_{-}]=-2K_{3}.
\end{equation}
We note that this (unitary) representation is necessarily infinite 
dimensional. 
The Fock space on which $\{ K_{+}, K_{-}, K_{3} \}$ act is 
$\calh_K \equiv \{\ket{K,n} \vert n\in\futon\cup\{0\} \}$ and 
whose actions are
\begin{eqnarray}
  \label{eq:2-2-4}
 K_{+} \ket{K,n} &=& \sqrt{(n+1)(2K+n)}\ket{K,n+1}, \quad 
 K_{-} \ket{K,n}  = \sqrt{n(2K+n-1)}\ket{K,n-1},   \nonumber  \\
 K_{3} \ket{K,n} &=& (K+n)\ket{K,n}, 
\end{eqnarray}
where $\ket{K,0}$ is a normalized vacuum ($K_{-}\ket{K,0}=0$ and 
$\langle K,0|K,0 \rangle =1$). We have written $\ket{K,0}$ instead 
of $\ket{0}$  to emphasize the spin $K$ representation, see \cite{FKSF1}. 
From (\ref{eq:2-2-4}), states $\ket{K,n}$ are given by 
\begin{equation}
  \label{eq:2-2-5}
 \ket{K,n} =\frac{(K_{+})^n}{\sqrt{n!(2K)_n}}\ket{K,0} ,
\end{equation}
where $(a)_n$ is the Pochammer's notation \ 
$
 (a)_n \equiv  a(a+1) \cdots (a+n-1).
$
These states satisfy the orthogonality and completeness conditions 
\begin{equation}
  \label{eq:2-2-7}
  \langle K,m \vert K,n \rangle =\delta_{mn}, 
 \quad \sum_{n=0}^{\infty}\ket{K,n}\bra{K,n}\ = \mathbf{1}_K.
\end{equation}
Now let us consider a generalized version of coherent states : 

\noindent{\bfseries Definition}\quad We call a state  
\begin{equation}
   \label{eq:2-2-8}
 \ket{z} = V(z)\ket{K,0} \equiv  
 \mbox{e}^{zK_{+} - \bar{z}K_{-}} \ket{K,0}  
  \quad \mbox{for} \quad z \in \fukuso.
\end{equation}
the generalized coherent state (or the coherent state of Perelomov's 
type based on $su(1,1)$ in our terminology).

Here let us construct an example of this representation. 
First we set
\begin{equation}
  \label{eq:2-2-12}
  K_{+}\equiv{1\over2}\left( a^{\dagger}\right)^2\ ,\
  K_{-}\equiv{1\over2}a^2\ ,\
  K_{3}\equiv{1\over2}\left( a^{\dagger}a+{1\over2}\right)\ ,
\end{equation}
then it is easy to check that these satisfy the commutation relations 
(\ref{eq:2-2-3}). 
That is, the set $\{K_{+},K_{-},K_{3}\}$ gives a unitary representation of 
$su(1,1)$ with spin $K = 1/4\ \mbox{and}\ 3/4$, \cite{AP}. 
Now we call an operator 
\begin{equation}
  \label{eq:2-2-14}
   S(z) = \mbox{e}^{\frac{1}{2}\{z(a^{\dagger})^2 - \bar{z}a^2\}}
   \quad \mbox{for} \quad z \in \fukuso 
\end{equation}
the squeezed operator, see \cite{AP} or \cite{WPS}.

\subsection{Generalized Coherent Operator Based on $su(2)$}
Let us state generalized coherent operators and states based on $su(2)$.

We consider a spin $J\ (> 0)$ representation of $su(2) 
\subset sl(2,\fukuso)$ and set its generators 
$\{ J_{+}, J_{-}, J_{3} \}\ ((J_{+})^{\dagger} = J_{-})$, 
\begin{equation}
  \label{eq:2-3-3}
 [J_{3}, J_{+}]=J_{+}, \quad [J_{3}, J_{-}]=-J_{-}, 
 \quad [J_{+}, J_{-}]=2J_{3}.
\end{equation}
We note that this (unitary) representation is necessarily finite 
dimensional. 
The Fock space on which $\{ J_{+}, J_{-}, J_{3} \}$ act is 
$\calh_{J} \equiv \{\ket{J,n} \vert 0 \le n \le 2J \}$ and 
whose actions are
\begin{eqnarray}
  \label{eq:2-3-4}
 J_{+} \ket{J,n} &=& \sqrt{(n+1)(2J-n)}\ket{J,n+1}, \quad 
 J_{-} \ket{J,n}  =  \sqrt{n(2J-n+1)}\ket{J,n-1},   \nonumber  \\
 J_{3} \ket{J,n} &=& (-J+n)\ket{J,n},
\end{eqnarray}
where $\ket{J,0}$ is a normalized vacuum ($J_{-}\ket{J,0}=0$ and 
$\langle J,0|J,0 \rangle =1$). We have written $\ket{J,0}$ instead 
of $\ket{0}$ to emphasize the spin $J$ representation, see \cite{FKSF1}. 
From (\ref{eq:2-3-4}), states $\ket{J,n}$ are given by 
\begin{equation}
  \label{eq:2-3-5}
 \ket{J,n} =\frac{(J_{+})^n}{\sqrt{n!{}_{2J}P_n}}\ket{J,0}.
\end{equation}
These states satisfy the orthogonality and completeness conditions 
\begin{equation}
  \label{eq:2-3-6}
  \langle J,m \vert J,n \rangle =\delta_{mn}, 
  \quad \sum_{n=0}^{2J}\ket{J,n}\bra{J,n}\ = \mathbf{1}_{J}.
\end{equation}
Now let us consider a generalized version of coherent states : 

\noindent{\bfseries Definition}\quad We call a state  
\begin{equation}
   \label{eq:2-3-7}
 \ket{z} = W(z)\ket{J,0} \equiv 
 \mbox{e}^{zJ_{+} - \bar{z}J_{-}} \ket{J,0}  
  \quad \mbox{for} \quad z \in \fukuso.
\end{equation}
the generalized coherent state (or the coherent state of Perelomov's 
type based on $su(2)$ in our terminology).

\subsection{Schwinger's Boson Method}
Here let us construct the spin $K$ and $J$ representations by making 
use of Schwinger's boson method.

Next we consider the system of two-harmonic oscillators. If we set
\begin{equation}
  \label{eq:2-4-1}
  a_1 = a \otimes 1,\  {a_1}^{\dagger} = a^{\dagger} \otimes 1;\ 
  a_2 = 1 \otimes a,\  {a_2}^{\dagger} = 1 \otimes a^{\dagger},
\end{equation}
then it is easy to see 
\begin{equation}
  \label{eq:2-4-2}
 [a_i, a_j] = [{a_i}^{\dagger}, {a_j}^{\dagger}] = 0,\ 
 [a_i, {a_j}^{\dagger}] = \delta_{ij}, \quad i, j = 1, 2. 
\end{equation}
We also denote by $N_{i} = {a_i}^{\dagger}a_i$ number operators.

Now we can construct representation of Lie algebras $su(2)$ and $su(1,1)$ 
making use of Schwinger's boson method, see \cite{FKSF1}, \cite{FKSF2}. 
Namely if we set 
\begin{eqnarray}
  \label{eq:2-4-3-1}
  su(2) &:&\quad
     J_+ = {a_1}^{\dagger}a_2,\ J_- = {a_2}^{\dagger}a_1,\ 
     J_3 = {1\over2}\left({a_1}^{\dagger}a_1 - {a_2}^{\dagger}a_2\right), \\
  \label{eq:2-4-3-2}
  su(1,1) &:&\quad
     K_+ = {a_1}^{\dagger}{a_2}^{\dagger},\ K_- = a_2 a_1,\ 
     K_3 = {1\over2}\left({a_1}^{\dagger}a_1 + {a_2}^{\dagger}a_2  + 1\right),
\end{eqnarray}
then these satisfy the commutation relations (\ref{eq:2-2-3}) and 
(\ref{eq:2-3-3}) respectively. 

\section{Matrix Elements of Coherent and Generalized Coherent Operators 
$\cdots$ \cite{KF14} }

\subsection{Matrix Elements of Coherent Operator}

We list matrix elements of coherent operators $U(z)$. 

\noindent{\bfseries The Matrix Elements}\quad The matrix elements of 
$U(z)$  are :
\begin{eqnarray}
   \label{eq:3-1-1-1}
 &&(\mbox{i})\quad n \le m \quad 
   \bra{n}U(z)\ket{m} = \mbox{e}^{-\frac{1}{2}\zetta^2}\sqrt{\frac{n!}{m!}}
                 (-\bar{z})^{m-n}{L_n}^{(m-n)}(\zetta^2), \\
   \label{eq:3-1-1-2}
 &&(\mbox{ii})\quad n \geq m \quad 
   \bra{n}U(z)\ket{m} = \mbox{e}^{-\frac{1}{2}\zetta^2}\sqrt{\frac{m!}{n!}}
                 z^{n-m}{L_m}^{(n-m)}(\zetta^2),
\end{eqnarray}
where ${L_n}^{(\alpha)}$ is the associated Laguerre's polynomial defined by 
\begin{equation}
   \label{eq:3-1-2}
 {L_k}^{(\alpha)}(x)=\sum_{j=0}^{k}(-1)^j {{k+\alpha}\choose{k-j}}
                  \frac{x^j}{j!}. 
\end{equation}
In particular $L_{k}\equiv {L_k}^{(0)}$ is the usual Laguerre's polynomial 
and these are related to diagonal elements of $U(z)$.

\subsection{Matrix Elements of Coherent Operator Based on $su(1,1)$}

We list matrix elements of $V(z)$  coherent operators based on $su(1,1)$. 
In this case it is always $2K > 1$ ($2K=1$ under some regularization). 

\noindent{\bfseries The Matrix Elements}\quad The matrix elements of 
$V(z)$ are :
\begin{eqnarray}
   \label{eq:3-2-1-1}
  (\mbox{i})\quad n \le m \quad 
   && \bra{K,n}V(z)\ket{K,m}= 
    \sqrt{\frac{n!m!}{(2K)_n(2K)_m}}
    {(-\bar{\kappa})^{m-n}}(1+\kappazetta^2)^{-K-\frac{n+m}{2}}
    \ \times   \nonumber \\
   &&\sum_{j=0}^{n}(-1)^{n-j}\frac{\Gamma(2K+m+n-j)}{\Gamma(2K)(m-j)!
     (n-j)!j!}(1+\kappazetta^2)^j(\kappazetta^2)^{n-j}, \\ 
   \label{eq:3-2-1-2}
  (\mbox{ii})\quad n \geq m \quad    
    && \bra{K,n}V(z)\ket{K,m}=
    \sqrt{\frac{n!m!}{(2K)_n(2K)_m}}
    {\kappa^{n-m}}(1+\kappazetta^2)^{-K-\frac{n+m}{2}}
    \ \times   \nonumber \\
   &&\sum_{j=0}^{m}(-1)^{m-j}\frac{\Gamma(2K+m+n-j)}{\Gamma(2K)(m-j)!
     (n-j)!j!}(1+\kappazetta^2)^j(\kappazetta^2)^{m-j}, 
\end{eqnarray}
where 
\begin{equation}
   \label{eq:3-2-2}
\kappa \equiv \frac{\mbox{sinh}(\zetta)}{\zetta}z 
       ={\mbox{cosh}(\zetta)}\zeta.
\end{equation}

The author doesn't know whether or not the right hand sides of 
(\ref{eq:3-2-1-1}) and (\ref{eq:3-2-1-2}) could be written by 
making use of some special functions such as generalized Laguerre's 
functions in (\ref{eq:3-1-2}). Therefore we set temporarily 
\begin{equation}
{F_{m}}^{(n-m)}(x:2K)=
     \sum_{j=0}^{m}(-1)^{m-j}\frac{\Gamma(2K+m+n-j)}{\Gamma(2K)(m-j)!
     (n-j)!j!}(1+x)^j{x}^{m-j} 
\end{equation}
and ${F_{m}}^{(0)}(x;2K)=F_{m}(x;2K)$.

\subsection{Matrix Elements of Coherent Operator Based on $su(2)$}

We list matrix elements of $W(z)$  coherent operators based on $su(2)$.
In this case it is always $2J \in \futon$.

\noindent{\bfseries  Matrix Elements}\quad The matrix elements of 
$W(z)$ are :
\begin{eqnarray}
   \label{eq:3-3-1-1}
  (\mbox{i})\quad n \le m \quad 
   && \bra{J,n}W(z)\ket{J,m}= 
    \sqrt{\frac{n!m!}{{}_{2J}P_n {}_{2J}P_m}}
    (-\bar{\kappa})^{m-n}(1-\kappazetta^2)^{J-\frac{n+m}{2}}
    \ \times   \nonumber \\
   &&\sum_{j=0}^{n}{}_{*}(-1)^{n-j}\frac{(2J)!}{(2J-m-n+j)!(m-j)!(n-j)!j!}
     (1-\kappazetta^2)^{j}(\kappazetta^2)^{n-j},\quad    \\ 
   \label{eq:3-3-1-2}
  (\mbox{ii})\quad n \geq m \quad    
    && \bra{J,n}W(z)\ket{J,m}=
    \sqrt{\frac{n!m!}{{}_{2J}P_n {}_{2J}P_m}}
    \kappa^{n-m}(1-\kappazetta^2)^{J-\frac{n+m}{2}}
    \ \times   \nonumber \\
   &&\sum_{j=0}^{m}{}_{*}(-1)^{m-j}\frac{(2J)!}{(2J-m-n+j)!(m-j)!(n-j)!j!}
     (1-\kappazetta^2)^{j}(\kappazetta^2)^{m-j},\quad  
\end{eqnarray}
where 
\begin{equation}
   \label{eq:3-3-2}
\kappa \equiv \frac{\mbox{sin}(\zetta)}{\zetta}z 
       ={\mbox{cos}(\zetta)}\eta.
\end{equation}
Here $\sum{}_{*}$ means a summation over $j$ satisfying $2J-m-n+j \geq 0$. 

The author doesn't know whether or not the right hand sides of 
(\ref{eq:3-3-1-1}) and (\ref{eq:3-3-1-2}) could be written 
by making use of some special functions. We set temporarily
\begin{equation}
{F_{m}}^{(n-m)}(x:2J)=
     \sum_{j=0}^{m}{}_{*}(-1)^{m-j}\frac{(2J)!}{(2J-m-n+j)!(m-j)!(n-j)!j!}
     (1-x)^{j}{x}^{m-j}
\end{equation}
and ${F_{m}}^{(0)}(x;2J)=F_{m}(x;2J)$.

\section{N--Level System in the Strong Coupling Region}

Let us make a short review of \cite{KF15} which is a generalization of 
\cite{MFr}. 

In \cite{MFr} Frasca dealt with a model that describes a 2--level atom 
interacting with a single radiation mode and developed some method 
to calculate Rabi frequencies in the strong coupling regime. 
We have generalized the model and method, and showed in \cite{KF15} 
that Rabi frequencies in our extended model were given by matrix elements 
of generalized coherent operators (\cite{KF14}) 
under the rotating--wave approximation. 

Let $\{\sigma_{1}, \sigma_{2}, \sigma_{3}\}$ be Pauli matrices and 
${\bf 1}_2$ a unit matrix : 
\begin{equation}
\sigma_{1} = 
\left(
  \begin{array}{cc}
    0& 1 \\
    1& 0
  \end{array}
\right), \quad 
\sigma_{2} = 
\left(
  \begin{array}{cc}
    0& -i \\
    i& 0
  \end{array}
\right), \quad 
\sigma_{3} = 
\left(
  \begin{array}{cc}
    1& 0 \\
    0& -1
  \end{array}
\right), \quad
{\bf 1}_2 = 
\left(
  \begin{array}{cc}
    1& 0 \\
    0& 1
  \end{array}
\right).
\end{equation}
We note here that 
\begin{equation}
\sigma_{1}=W\sigma_{3}W^{-1}
\end{equation}
where $W$ is given by 
\begin{equation}
\label{eq:Walsh--Hadamard matrix}
W=\frac{1}{\sqrt{2}}
\left(
  \begin{array}{cc}
    1& 1\\
    1& -1
  \end{array}
\right)
=W^{-1}
\end{equation}
and is known as the Walsh--Hadamard matrix. 

\par \noindent 
The Hamiltonians adopted in \cite{KF15} are 
\begin{eqnarray}
\label{eq:three-Hamiltonians}
\mbox{(N)}\qquad H_{N}&=&\omega {\bf 1}_{2}\otimes a^{\dagger}a + 
\frac{\Delta}{2}\sigma_{3}\otimes {\bf 1} +
g\sigma_{1}\otimes (a^{\dagger}+a), \nonumber \\
\mbox{(K)}\qquad H_{K}&=&\omega {\bf 1}_{2}\otimes K_{3} + 
\frac{\Delta}{2}\sigma_{3}\otimes {\bf 1}_{K} +
g\sigma_{1}\otimes (K_{+}+K_{-}), \\
\mbox{(J)}\qquad H_{J}&=&\omega {\bf 1}_{2}\otimes J_{3} + 
\frac{\Delta}{2}\sigma_{3}\otimes {\bf 1}_{J} +
g\sigma_{1}\otimes (J_{+}+J_{-}), \nonumber 
\end{eqnarray}
where $\omega$ is the frequency of the radiation mode, $\Delta$ 
the separation  between the two levels of the atom, $g$ the coupling 
between the radiation field and the atom.

\par \noindent 
To deal with these three cases at the same time we set 
\begin{equation}
\{L_{+},L_{-},L_{3}\}=
\left\{
\begin{array}{ll}
\mbox{(N)}\qquad \{a^{\dagger},a,N\} \\
\mbox{(K)}\qquad \{K_{+},K_{-},K_{3}\} \\
\mbox{(J)}\qquad \ \{J_{+},J_{-},J_{3}\} 
\end{array}
\right.
\end{equation}
and 
\begin{equation}
H=H_{0}+V
=\omega {\bf 1}_{2}\otimes L_{3} + 
\frac{\Delta}{2}\sigma_{3}\otimes {\bf 1}_{L} +
g\sigma_{1}\otimes (L_{+}+L_{-})
\end{equation}
where we have written $H$ instead of $H_{L}$ for simplicity. 

(\ref{eq:three-Hamiltonians}) is based on an atom with 2--level. However  
an atom has in reality many (energy) levels, so it is natural to consider 
an atom with n--level. First of all let us make some mathematical 
preliminaries, see for example \cite{KF7}; Appendix. 

\par \noindent 
Let $\{\Sigma_{1}, \Sigma_{3}\}$ be matrices in $M(n,\fukuso)$ 
\begin{equation}
\label{Sigma-1}
\Sigma_{1}=
\left(
\begin{array}{ccccccc}
0&  &  &      &      &      & 1  \\
1& 0&  &      &      &      &    \\
 & 1& 0&      &      &      &    \\
 &  & 1& \cdot&      &      &    \\
 &  &  & \cdot& \cdot&      &    \\
 &  &  &      & \cdot& \cdot&    \\
 &  &  &      &      &    1 & 0
\end{array}
\right),      \qquad 
\label{Sigma-3}
\Sigma_{3}=
\left(
\begin{array}{ccccccc}
1&       &           &      &      &      &             \\
 & \sigma&           &      &      &      &             \\
 &       & {\sigma}^2&      &      &      &             \\
 &       &           & \cdot&      &      &             \\
 &       &           &      & \cdot&      &             \\
 &       &           &      &      & \cdot&             \\
 &       &           &      &      &      & {\sigma}^{n-1}
\end{array}
\right)
\end{equation}
where $\sigma$ is a primitive element satisfying ${\sigma}^{n}=1$ (
$\sigma=\mbox{e}^{\frac{2\pi \sqrt{-1}}{n}}$). 
We note that $\{\Sigma_{1}, \Sigma_{3}\}$ is a generalization of Pauli 
matrices $\{\sigma_{1}, \sigma_{3}\}$, but they are not hermitian. 
If we define a Vandermonde matrix $W$ based on $\sigma$ as 
\begin{eqnarray}
\label{eq:Large-double}
W&=&\frac{1}{\sqrt{n}}
\left(
\begin{array}{ccccccc}
1&        1&     1&   \cdot & \cdot  & \cdot & 1             \\
1& \sigma^{n-1}& \sigma^{2(n-1)}&  \cdot& \cdot& \cdot& \sigma^{(n-1)^2} \\
1& \sigma^{n-2}& \sigma^{2(n-2)}&  \cdot& \cdot& \cdot& \sigma^{(n-1)(n-2)} \\
\cdot&  \cdot &  \cdot  &     &      &      & \cdot  \\
\cdot&  \cdot  & \cdot &      &      &      &  \cdot  \\
1& \sigma^{2}& \sigma^{4}& \cdot & \cdot & \cdot & \sigma^{2(n-1)} \\
1& \sigma & \sigma^{2}& \cdot& \cdot& \cdot& \sigma^{n-1}
\end{array}
\right), \\
\label{eq:Large-double-dagger}
W^{\dagger}&=&\frac{1}{\sqrt{n}}
\left(
\begin{array}{ccccccc}
1&        1&     1&   \cdot & \cdot  & \cdot & 1             \\
1& \sigma& \sigma^{2}&  \cdot& \cdot& \cdot& \sigma^{n-1} \\
1& \sigma^{2}& \sigma^{4}&  \cdot& \cdot& \cdot& \sigma^{2(n-1)} \\
\cdot&  \cdot &  \cdot  &     &      &      & \cdot  \\
\cdot&  \cdot  & \cdot &      &      &      &  \cdot  \\
1& \sigma^{n-2}& \sigma^{2(n-2)}& \cdot& \cdot& \cdot& \sigma^{(n-1)(n-2)} \\
1&    \sigma^{n-1} & \sigma^{2(n-1)}& \cdot& \cdot& \cdot& \sigma^{(n-1)^2}
\end{array}
\right), 
\end{eqnarray}
then it is not difficult to see 
\begin{equation}
\label{eq:diagonalization}
\Sigma_{1}=W\Sigma_{3}W^{\dagger}=W\Sigma_{3}W^{-1}.
\end{equation}
For example, for $n=3$
\begin{eqnarray}
W\Sigma_{3}W^{\dagger}&=&\frac{1}{3}
\left(
\begin{array}{ccc}
1&          1&     1     \\
1& \sigma^{2}& \sigma    \\
1& \sigma    & \sigma^{2}
\end{array}
\right)
\left(
\begin{array}{ccc}
1&       &           \\
 & \sigma&           \\
 &       & {\sigma}^2 
\end{array}
\right)
\left(
\begin{array}{ccc}
1&        1&     1 \\
1& \sigma & \sigma^{2} \\
1& \sigma^{2}  & \sigma
\end{array}
\right)   \nonumber \\
&=&\frac{1}{3}
\left(
\begin{array}{ccc}
1& \sigma& \sigma^{2}\\
1&      1&  1        \\
1& \sigma^{2}& \sigma 
\end{array}
\right)
\left(
\begin{array}{ccc}
1&        1&     1 \\
1& \sigma& \sigma^{2} \\
1& \sigma^{2}  & \sigma
\end{array}
\right)  
=\frac{1}{3}
\left(
\begin{array}{ccc}
0&  0& 3\\
3&  0& 0\\
0&  3& 0
\end{array}
\right) 
=\Sigma_{1},  \nonumber
\end{eqnarray}
where we have used that $\sigma^{3}=1,\ \bar{\sigma}=\sigma^{2}\ \mbox{and}\ 
1+\sigma+\sigma^{2}=0$. 

\par \noindent 
That is, $\Sigma_{1}$ can be diagonalized by making use of $W$. 

\par \noindent
A comment is in order. Since $W$ corresponds to the Walsh--Hadamard matrix
(\ref{eq:Walsh--Hadamard matrix}), so it may be possible to call $W$ 
the generalized Walsh--Hadamard matrix. 

\par \noindent 
From this we know all eigenvalues and eigenstates of $\Sigma_{1}$ 
to become 
\begin{eqnarray}
\mbox{Eigenvalues}&=&\{1, \sigma, \cdots, \sigma^{j}, \cdots, \sigma^{n-1}\} \\
\mbox{Eigenstates}&=&\{\ket{1}, \ket{\sigma}, \cdots, \ket{\sigma^j}, \cdots, 
\ket{\sigma^{n-1}}\}
\end{eqnarray}
where@
\[
\ket{\sigma^j}=\frac{1}{\sqrt{n}}
\left(
\begin{array}{c}
1 \\
\sigma^{(n-1)j} \\
\sigma^{(n-2)j} \\
\cdot \\
\cdot \\
\sigma^{2j}\\
\sigma^{j}
\end{array}
\right)\quad \mbox{for}\quad 0 \leq j \leq n-1 .
\]
Therefore $\Sigma_{1}$ can be written as (a spectral decomposition) 
\begin{equation}
\label{eq:spectral-decomposition}
\Sigma_{1}=\sum_{j=0}^{n-1}\sigma^{j}\ket{\sigma^{j}}\bra{\sigma^{j}} 
\quad \mbox{and}\quad 
{\bf 1}_{n}=\sum_{j=0}^{n-1}\ket{\sigma^{j}}\bra{\sigma^{j}}. 
\end{equation}

The usual model is based on an atom with 2--level and we 
want to consider the case of an atom with n--level. Therefore it is very 
natural for us to generalize the model by making use of 
$\{\Sigma_{1}, \Sigma_{3}\}$ in place of $\{\sigma_{1},\sigma_{3}\}$, 
but then we meet some trouble. One reason is $\Sigma_{1}$ and $\Sigma_{3}$ 
are {\bf not} hermitian. The Hamiltonians must be hermitian in general. 

After some trial and error we reach the following : 
\begin{equation}
\label{eq:Starting-Hamiltonians}
H=H_{0}+V
=\omega {\bf 1}_{n}\otimes L_{3} + 
\frac{\bar{\Delta}}{2}\Sigma_{3}\otimes {\bf 1}_{L} + 
\frac{\Delta}{2}{\Sigma_{3}}^{\dagger}\otimes {\bf 1}_{L} + 
g(\Sigma_{1}\otimes L_{+} + {\Sigma_{1}}^{\dagger}\otimes L_{-})
\end{equation}
where $\Delta$ is a complex constant (we cannot prevent a {\bf complex} 
constant). We believe that our extension is very natural except this 
point.

\par \noindent 
We cannot solve these simple models completely (maybe non--integrable), 
but we have found these models have a rich structure. 

\par \noindent 
For these models we usually have two perturbation approaches : 

\noindent{\bfseries Weak Coupling Regime} ($0 < g \ll |\Delta|$)\quad 
\begin{equation}
\label{eq:Weak-Coupling-Regime}
H_{0}=
\omega {\bf 1}_{n}\otimes L_{3} + 
\frac{\bar{\Delta}}{2}\Sigma_{3}\otimes {\bf 1}_{L} + 
\frac{\Delta}{2}{\Sigma_{3}}^{\dagger}\otimes {\bf 1}_{L}, \quad 
V=g(\Sigma_{1}\otimes L_{+}+ {\Sigma_{1}}^{\dagger}\otimes L_{-}). 
\end{equation}

\noindent{\bfseries Strong Coupling Regime} ($0 < |\Delta| \ll g$)\quad 
\begin{equation}
\label{eq:Strong-Coupling-Regime}
H_{0}=
\omega {\bf 1}_{n}\otimes L_{3} + 
g(\Sigma_{1}\otimes L_{+}+ {\Sigma_{1}}^{\dagger}\otimes L_{-}), \quad 
V=\frac{\bar{\Delta}}{2}\Sigma_{3}\otimes {\bf 1}_{L} + 
\frac{\Delta}{2}{\Sigma_{3}}^{\dagger}\otimes {\bf 1}_{L}. 
\end{equation}

\par \noindent 
In this paper we are not interested in the weak coupling regime, so  
in the following we focus our attention on the strong coupling regime. 
We would like to generalize the result in \cite{KF15}. 

\par \noindent 
Then we have from (\ref{eq:spectral-decomposition}) and 
(\ref{eq:Strong-Coupling-Regime}) 
\begin{eqnarray}
\label{eq:Basic-Hamiltonian}
H_{0}&=&
\sum_{j=0}^{n-1}\ket{\sigma^{j}}\bra{\sigma^{j}}\otimes 
\{\omega L_{3} + g(\sigma^{j}L_{+}+\bar{\sigma}^{j}L_{-})\}  \nonumber \\
&=&\sum_{j=0}^{n-1}\ket{\sigma^{j}}\bra{\sigma^{j}}\otimes 
\left\{
\mbox{e}^{-\frac{1}{2}(z_{j}L_{+}-\bar{z}_{j}L_{-})}
\left(\Omega L_{3}\right)
\mbox{e}^{\frac{1}{2}(z_{j}L_{+}-\bar{z}_{j}L_{-})}
\right\}    \nonumber \\
&=&\sum_{j=0}^{n-1}
\left(
\ket{\sigma^{j}}\otimes \mbox{e}^{-\frac{1}{2}(z_{j}L_{+}-\bar{z}_{j}L_{-})}
\right)
\left(\Omega L_{3}\right)
\left(
\bra{\sigma^{j}}\otimes \mbox{e}^{\frac{1}{2}(z_{j}L_{+}-\bar{z}_{j}L_{-})}
\right) 
\end{eqnarray}
where we have used the following 

\noindent{\bfseries Key Formulas}
\begin{eqnarray}
\label{eq:N-formula}
&&(N)\quad 
\omega a^{\dagger}a + g(\sigma^{j} a^{\dagger}+\bar{\sigma}^{j}a)
=\Omega\ \mbox{e}^{-\frac{1}{2}(z_{j} a^{\dagger}-\bar{z}_{j}a)}
\left(N-\frac{g^2}{\omega^2}\right)
\mbox{e}^{\frac{1}{2}(z_{j} a^{\dagger}-\bar{z}_{j}a)} \nonumber \\
&&\qquad  \quad \mbox{where}\quad 
\Omega=\omega, \quad  z_{j}=(2g/\omega)\sigma^{j}, \\
\label{eq:K-formula}
&&(K)\quad 
\omega K_{3} + g(\sigma^{j} K_{+}+\bar{\sigma}^{j}K_{-})
=\Omega\ \mbox{e}^{-\frac{1}{2}(z_{j}K_{+}-\bar{z}_{j}K_{-})} K_{3}\ 
\mbox{e}^{\frac{1}{2}(z_{j}K_{+}-\bar{z}_{j}K_{-})} \nonumber \\
&&\qquad  \quad \mbox{where}\quad 
\Omega=\omega \sqrt{1-(2g/\omega)^{2}}, \quad 
z_{j}=\mbox{tanh}^{-1}(2g/\omega)\sigma^{j}, \\
\label{eq:J-formula}
&&(J)\quad
\omega J_{3} + g(\sigma^{j} J_{+}+\bar{\sigma}^{j}J_{-})
=\Omega\ \mbox{e}^{-\frac{1}{2}(z_{j}J_{+}-\bar{z}_{j}J_{-})} J_{3}\ 
\mbox{e}^{\frac{1}{2}(z_{j}J_{+}-\bar{z}_{j}J_{-})} \nonumber \\
&&\qquad  \quad \mbox{where}\quad 
\Omega=\omega \sqrt{1+(2g/\omega)^{2}}, \quad 
z_{j}=\mbox{tan}^{-1}(2g/\omega)\sigma^{j}. 
\end{eqnarray}
These formulas are a slight generalization of corresponding ones 
in \cite{KF15}. We leave the proof to the readers. 

\par \noindent 
That is, we could diagonalize the Hamiltonian $H_{0}$. This is $n$--fold 
degenerate and its eigenvalues and eigenvectors are given respectively 
\begin{equation}
\label{eq:Eigenvalues-Eigenvectors}
(\mbox{Eigenvalues},\mbox{Eigenvectors})=
\left\{
\begin{array}{ll}
(N)\quad \Omega (-\frac{g^2}{\omega^2}+m),\quad 
\ket{\sigma^{j}}\otimes 
\mbox{e}^{-\frac{1}{2}(z_{j}a^{\dagger}-\bar{z}_{j}a)}\ket{m} \\
(K)\quad \Omega (K+m),\quad 
\ket{\sigma^{j}}\otimes 
\mbox{e}^{-\frac{1}{2}(z_{j}K_{+}-\bar{z}_{j}K_{-})}\ket{K,m} \\
(J)\quad \Omega (-J+m),\quad 
\ket{\sigma^{j}}\otimes 
\mbox{e}^{-\frac{1}{2}(z_{j}J_{+}-\bar{z}_{j}J_{-})}\ket{J,m} \\
\end{array}
\right.
\end{equation}
for $j=0,1,\cdots,n-1$ and $m \in \futon \cup \{0\}$. 
For the latter convenience we set 
\begin{equation}
\label{eq:eigenvalues-eigenvectors}
\mbox{Eigenvalues}=\{E_{m}\}, \quad 
\mbox{Eigenvectors}=\{\ket{\{\sigma^{j}, m\}}\}. 
\end{equation}
Then (\ref{eq:Basic-Hamiltonian}) can be written as 
\begin{equation}
\label{eq:Basic-Hamiltonian-2}
H_{0}
=\sum_{j}\sum_{m}E_{m}\ket{\{\sigma^{j}, m\}}\bra{\{\sigma^{j}, m\}}.
\end{equation}

\par \vspace{5mm} 
Next we would like to solve the following Schr{\"o}dinger equation : 
\begin{equation}
\label{eq:full-equation}
i\frac{d}{dt}\Psi=H\Psi=
\left(
H_{0}+\frac{\bar{\Delta}}{2}\Sigma_{3}\otimes {\bf 1}_{L}+
\frac{\Delta}{2}{\Sigma_{3}}^{\dagger}\otimes {\bf 1}_{L}
\right)\Psi, 
\end{equation}
where we have set $\hbar=1$ for simplicity. 
To solve this equation we appeal to the {\bf method of constant variation}. 
First let us solve 
\begin{equation}
\label{eq:partial-equation}
i\frac{d}{dt}\Psi=H_{0}\Psi, 
\end{equation}
which general solution is given by 
\begin{equation}
\label{eq:partial-solution}
  \Psi(t)=U_{0}(t)\Psi_{0}=\mbox{e}^{-itH_{0}}\Psi_{0}
\end{equation}
where $\Psi_{0}$ is a constant state. It is easy to see from 
(\ref{eq:Basic-Hamiltonian-2})
\begin{equation}
\label{eq:Basic-Unitary}
U_{0}(t)=\mbox{e}^{-itH_{0}}
=\sum_{j}\sum_{m}\mbox{e}^{-itE_{m}}
\ket{\{\sigma^{j}, m\}}\bra{\{\sigma^{j}, m\}}.
\end{equation}

The method of constant variation goes as follows. Changing like \ \ 
$
\Psi_{0} \longrightarrow \Psi_{0}(t),
$ \ \ 
we insert (\ref{eq:partial-solution}) into (\ref{eq:full-equation}). 
After some algebra we obtain 
\begin{equation}
\label{eq:sub-equation}
i\frac{d}{dt}\Psi_{0}
=\left\{
\frac{\Delta}{2}
{U_0}^{\dagger}\left({\Sigma_{3}}^{\dagger}\otimes {\bf 1}_{L}\right){U_0}\ 
+ \ 
\frac{\bar{\Delta}}{2}
{U_0}^{\dagger}\left(\Sigma_{3}\otimes {\bf 1}_{L}\right){U_0}
\right\}\Psi_{0} 
\equiv (H_{F}+ {H_{F}}^{\dagger})\Psi_{0}.
\end{equation}
Now let us calculate $H_{F}$. From (\ref{eq:Basic-Unitary}) and 
(\ref{eq:Eigenvalues-Eigenvectors}) 
\begin{eqnarray}
H_{F}&=&\frac{\Delta}{2}
\sum_{j,k}\sum_{m,r}
\mbox{e}^{it(E_m-E_r)} 
\bra{\{\sigma^{j},m\}}({\Sigma_{3}}^{\dagger}\otimes {\bf 1}_{L})
\ket{\{\sigma^{k},r\}} \ 
\ket{\{\sigma^{j}, m\}}\bra{\{\sigma^{k}, r\}} \nonumber \\
&=&\frac{\Delta}{2}
\sum_{j}\sum_{m, r}
\mbox{e}^{it\Omega(m-r)} 
\braa{m}\mbox{e}^{\frac{1}{2}
\left\{(z_{j}-z_{j-1})L_{+}-(\bar{z}_{j}-\bar{z}_{j-1})L_{-}\right\}
}\kett{r} \ 
\ket{\{\sigma^{j}, m\}}\bra{\{\sigma^{j-1}, r\}} 
\end{eqnarray}
where we have used the relation 
$\bra{\sigma^{j}}{\Sigma_{3}}^{\dagger}=\bra{\sigma^{j-1}}$. \ 
Remind that $\kett{m}$ is respectively 
\begin{equation}
\kett{m}=
\left\{
\begin{array}{ll}
(N)\qquad \ket{m} \\
(K)\qquad \ket{K,m} \\
(J)\qquad \ \ket{J,m}
\end{array}
\right.
\end{equation}

\par \noindent 
In this stage we meet {\bf matrix elements of the coherent and generalized 
coherent operators} $\mbox{e}^{zL_{+}-\bar{z}L_{-}}$ in section 3. \quad 
Here we divide $H_{F}$ into two parts
\[
H_{F}={H_{F}}^{'}+{H_{F}}^{''}
\]
where 
\begin{eqnarray}
\label{eq:Second-Hamiltonian-1}
{H_{F}}^{'}&=&\frac{\Delta}{2}\sum_{j}\sum_{m}
\braa{m}\mbox{e}^{ \frac{1}{2}
\left\{(z_{j}-z_{j-1})L_{+}-(\bar{z}_{j}-\bar{z}_{j-1})L_{-}\right\}
}\kett{m} \ 
\ket{\{\sigma^{j},m\}}\bra{\{\sigma^{j-1},m\}}, \\
\label{eq:Second-Hamiltonian-2}
{H_{F}}^{''}&=&\frac{\Delta}{2}\sum_{j}
\sum_{\stackrel{\scriptstyle m,r}{m\ne r}}
\mbox{e}^{it\Omega(m-r)} 
\braa{m}\mbox{e}^{ \frac{1}{2}
\left\{(z_{j}-z_{j-1})L_{+}-(\bar{z}_{j}-\bar{z}_{j-1})L_{-}\right\}
}\kett{r} \ 
\ket{\{\sigma^{j}, m\}}\bra{\{\sigma^{j-1}, r\}}.
\end{eqnarray}
Here let us deal with ${H_{F}}^{'}$. \  
By the formulas in section 3 it is easy to see 
\begin{eqnarray}
&&\braa{m}
\mbox{e}^{ \frac{1}{2}
\left\{(z_{0}-z_{-1})L_{+}-(\bar{z}_{0}-\bar{z}_{-1})L_{-}\right\} }
\kett{m} 
=
\braa{m}
\mbox{e}^{ \frac{1}{2}
\left\{(z_{1}-z_{0})L_{+}-(\bar{z}_{1}-\bar{z}_{0})L_{-}\right\} }
\kett{m} 
=
\cdots   \nonumber \\
=
&&\braa{m}
\mbox{e}^{ \frac{1}{2}
\left\{(z_{n-1}-z_{n-2})L_{+}-(\bar{z}_{n-1}-\bar{z}_{n-2})L_{-}\right\} }
\kett{m} 
\end{eqnarray}
because by {\bf Key Formulas} ((\ref{eq:N-formula}), (\ref{eq:K-formula}), 
(\ref{eq:J-formula}))
\begin{eqnarray}
&&z_{j}-z_{j-1}=C\sigma^{j-1}(\sigma-1)=C\sigma^{j-1}\sigma^{1/2}
(\sigma^{1/2}-\sigma^{-1/2})=2C\mbox{sin}(\pi/n)i\sigma^{j-1/2},  \\
&&|z_{j}-z_{j-1}|=2C\mbox{sin}(\pi/n) 
\quad \mbox{for} \quad j=0,1,\cdots, n-1   \nonumber
\end{eqnarray}
where $C$ is a constant defined by 
\begin{equation}
\label{eq:Constants}
C=
\left\{
\begin{array}{ll}
(N)\qquad 2g/\omega \\
(K)\qquad \mbox{tanh}^{-1}(2g/\omega) \\
(J)\qquad \ \mbox{tan}^{-1}(2g/\omega)
\end{array}
\right.
\end{equation}

\par \noindent 
Therefore ${H_{F}}^{'}$ can be written as 
\[
{H_{F}}^{'}=\frac{\Delta}{2}\sum_{m}
\braa{m}\mbox{e}^{ \frac{1}{2}
\left\{(z_{0}-z_{-1})L_{+}-(\bar{z}_{0}-\bar{z}_{-1})L_{-}\right\}
}\kett{m}
\left\{
\sum_{j}\ket{\{\sigma^{j},m\}}\bra{\{\sigma^{j-1},m\}}
\right\}.
\]
Let us diagonalize the last term. For that we set for simplicity 
\[
\Lambda_{m}
=\left(
\ket{\{1,m\}},\ket{\{\sigma^{1},m\}},\cdots, \ket{\{\sigma^{n-1},m\}}
\right).
\]
Then it is easy to see from (\ref{eq:diagonalization}) 
\[
\sum_{j=0}^{n-1}\ket{\{\sigma^{j},m\}}\bra{\{\sigma^{j-1},m\}}
=
\Lambda_{m} \Sigma_{1} \Lambda_{m}^{\dagger}
=
\Lambda_{m} W \Sigma_{3} W^{\dagger} \Lambda_{m}^{\dagger}
\equiv 
{\widetilde \Lambda}_{m} \Sigma_{3}{\widetilde \Lambda}_{m}^{\dagger}. 
\]
Here if we introduce a new basis 
\[
\Lambda_{m} W={\widetilde \Lambda}_{m}=
\left(
\ket{\{1,\psi_{m}\}}, \ket{\{\sigma,\psi_{m}\}}, \cdots, 
\ket{\{\sigma^{n-1},\psi_{m}\}}
\right), 
\]
where 
\begin{eqnarray}
\label{eq:multi-cat-states}
&&\ket{\{\sigma^{j},\psi_{m}\}}
\equiv ({\widetilde \Lambda}_{m})_{j}
=\sum_{k=0}^{n-1}(\Lambda_{m})_{k}W_{kj}    \nonumber \\
&&=\frac{1}{\sqrt{n}}
\left(
\ket{\{1,m\}}+\sigma^{j(n-1)}\ket{\{\sigma^{1},m\}}+
\sigma^{j(n-2)}\ket{\{\sigma^{2},m\}}+\cdots 
+\sigma^{j}\ket{\{\sigma^{n-1},m\}}
\right)
\end{eqnarray}
then we obtain 
\begin{equation}
\label{eq:Second-Hamiltonian-1-1}
{H_{F}}^{'}=\frac{\Delta}{2}\sum_{m}\sum_{j}
\braa{m}\mbox{e}^{ \frac{1}{2}
\left\{(z_{0}-z_{-1})L_{+}-(\bar{z}_{0}-\bar{z}_{-1})L_{-}\right\}
}\kett{m}
\sigma^{j}\ket{\{\sigma^{j},\psi_{m}\}}\bra{\{\sigma^{j},\psi_{m}\}}.
\end{equation}
This is just diagonal !  

\par \noindent 
A comment is in order. \ We would like to call the states 
(\ref{eq:multi-cat-states}) {\bf multi cat states of Schr{\" o}dinger} 
following from \cite{WPS} (because we are dealing with multi level systems). 

\par \vspace{5mm} 
Next let us deal with ${H_{F}}^{''}$.  By $\Lambda_{m}=
{{\widetilde \Lambda}_{m}}W^{\dagger}$ and (\ref{eq:Large-double}), 
(\ref{eq:Large-double-dagger}) we have 
\begin{eqnarray}
&&\ket{\{\sigma^{j},m\}}\bra{\{\sigma^{j-1},r\}}
=
(\Lambda_{m})_{j} ({\Lambda_{r}}^{\dagger})_{j-1}
=
({{\widetilde \Lambda}_{m}}W^{\dagger})_{j}
(W{{\widetilde \Lambda}_{r}}^{\dagger})_{j-1}
\nonumber \\
&&=
\sum_{k=0}^{n-1}\sum_{l=0}^{n-1} 
({\widetilde \Lambda}_{m})_{k}(W^{\dagger})_{kj}
W_{j-1,l}({{\widetilde \Lambda}_{r}}^{\dagger})_{l}
=
\sum_{k,l}(W^{\dagger})_{kj}W_{j-1,l}
\ket{\{\sigma^{k},\psi_{m}\}}\bra{\{\sigma^{l},\psi_{r}\}}
\nonumber \\
&&=
\frac{1}{n}\sum_{k,l}\sigma^{jk+(n-j+1)l}
\ket{\{\sigma^{k},\psi_{m}\}}\bra{\{\sigma^{l},\psi_{r}\}} 
\quad \mbox{or} \quad 
=
\frac{1}{n}\sum_{k,l}\sigma^{jk-(j-1)l}
\ket{\{\sigma^{k},\psi_{m}\}}\bra{\{\sigma^{l},\psi_{r}\}},
\nonumber 
\end{eqnarray}
so that we obtain
\begin{eqnarray}
\label{eq:Second-Hamiltonian-2-2}
{H_{F}}^{''}&=&\frac{\Delta}{2}
\sum_{\stackrel{\scriptstyle m,r}{m\ne r}}\sum_{k,l}
\mbox{e}^{it\Omega(m-r)} 
\left\{
\sum_{j=0}^{n-1}
\braa{m}\mbox{e}^{ \frac{1}{2}
\left\{(z_{j}-z_{j-1})L_{+}-(\bar{z}_{j}-\bar{z}_{j-1})L_{-}\right\}
}\kett{r} \ 
\right. \times \nonumber \\
&&
\left.
\frac{\sigma^{jk-(j-1)l}}{n}
\ket{\{\sigma^{k},\psi_{m}\}}\bra{\{\sigma^{l},\psi_{r}\}}
\right\}. 
\end{eqnarray}

\par \noindent 
Let us sum up the above result : 
\[
H_{F}={H_{F}}^{'}+{H_{F}}^{''},\quad 
{H_{F}}^{'}=(\ref{eq:Second-Hamiltonian-1-1}) \quad \mbox{and} \quad 
{H_{F}}^{''}=(\ref{eq:Second-Hamiltonian-2-2}). 
\]

Next we take the dagger of $H_{F}$ which is now an easy task.
\[
{H_{F}}^{\dagger}\equiv {\tilde{H}_{F}}^{'}+{\tilde{H}_{F}}^{''}
\]
where 
\begin{equation}
\label{eq:Second-Hamiltonian-1-1-dagger}
{\tilde{H}_{F}}^{'}=
\frac{\bar{\Delta}}{2}\sum_{m}\sum_{j}
\braa{m}\mbox{e}^{-\frac{1}{2}
\left\{(z_{0}-z_{-1})L_{+}-(\bar{z}_{0}-\bar{z}_{-1})L_{-}\right\}
}\kett{m}
\sigma^{-j}\ket{\{\sigma^{j},\psi_{m}\}}\bra{\{\sigma^{j},\psi_{m}\}},  
\end{equation}
and 
\begin{eqnarray}
\label{eq:Second-Hamiltonian-2-2-dagger}
{\tilde{H}_{F}}^{''}&=&\frac{\bar{\Delta}}{2}
\sum_{\stackrel{\scriptstyle m,r}{m\ne r}}\sum_{k,l}
\mbox{e}^{-it\Omega(m-r)} 
\left\{
\sum_{j=0}^{n-1}
\braa{r}\mbox{e}^{-\frac{1}{2}
\left\{(z_{j}-z_{j-1})L_{+}-(\bar{z}_{j}-\bar{z}_{j-1})L_{-}\right\}
}\kett{m} \ 
\right. \times \nonumber \\
&&\quad 
\left.
\frac{\bar{\sigma}^{jk-(j-1)l}}{n}
\ket{\{\sigma^{l},\psi_{r}\}}\bra{\{\sigma^{k},\psi_{m}\}}
\right\}
\nonumber \\
&=&\frac{\bar{\Delta}}{2}
\sum_{\stackrel{\scriptstyle m,r}{m\ne r}}\sum_{k,l}
\mbox{e}^{it\Omega(r-m)} 
\left\{
\sum_{j=0}^{n-1}
\braa{r}\mbox{e}^{-\frac{1}{2}
\left\{(z_{j}-z_{j-1})L_{+}-(\bar{z}_{j}-\bar{z}_{j-1})L_{-}\right\}
}\kett{m} \ 
\right. \times \nonumber \\
&&\quad 
\left.
\frac{{\sigma}^{(j-1)l-jk}}{n}
\ket{\{\sigma^{l},\psi_{r}\}}\bra{\{\sigma^{k},\psi_{m}\}}
\right\}
\nonumber \\
&=&\frac{\bar{\Delta}}{2}
\sum_{\stackrel{\scriptstyle m,r}{m\ne r}}\sum_{k,l}
\mbox{e}^{it\Omega(m-r)} 
\left\{
\sum_{j=0}^{n-1}
\braa{m}\mbox{e}^{-\frac{1}{2}
\left\{(z_{j}-z_{j-1})L_{+}-(\bar{z}_{j}-\bar{z}_{j-1})L_{-}\right\}
}\kett{r} \ 
\right. \times \nonumber \\
&&\quad 
\left.
\frac{{\sigma}^{(j-1)k-jl}}{n}
\ket{\{\sigma^{k},\psi_{m}\}}\bra{\{\sigma^{l},\psi_{r}\}}
\right\}.
\end{eqnarray}

\par \noindent 
Therefore 
\[
H_{F}+{H_{F}}^{\dagger}=
\left( {H_{F}}^{'}+{\tilde{H}_{F}}^{'} \right) +
\left( {H_{F}}^{''}+{\tilde{H}_{F}}^{''} \right),
\]
where 
\begin{eqnarray}
&&{H_{F}}^{'}+{\tilde{H}_{F}}^{'}
= 
\sum_{m}\sum_{j}{\Theta}_{m,j}
\ket{\{\sigma^{j},\psi_{m}\}}\bra{\{\sigma^{j},\psi_{m}\}},   \\
&&{\Theta}_{m,j}\equiv 
\frac{\Delta}{2}\sigma^{j}
\braa{m}\mbox{e}^{\frac{1}{2}
\left\{(z_{0}-z_{-1})L_{+}-(\bar{z}_{0}-\bar{z}_{-1})L_{-}\right\}
}\kett{m}
+
\frac{\bar{\Delta}}{2}\sigma^{-j}
\braa{m}\mbox{e}^{-\frac{1}{2}
\left\{(z_{0}-z_{-1})L_{+}-(\bar{z}_{0}-\bar{z}_{-1})L_{-}\right\}
}\kett{m}.   \nonumber \\
&&
\end{eqnarray}
Now for the latter use let us determine ${\Theta}_{m,j}$ : 
\begin{equation}
{\Theta}_{m,j}=
\left\{
\begin{array}{ll}
(N)\quad \frac{1}{2}
\left(\Delta \sigma^{j}+\bar{\Delta} \sigma^{-j} \right)
\mbox{exp}\left(-\frac{1}{2}|\kappa|^2 \right)
L_{m}\left(|\kappa|^2 \right) \\
\qquad \qquad \qquad \qquad \qquad \mbox{where}\quad 
\kappa=C\mbox{sin}(\pi/n)i,  \\
(K)\quad \frac{1}{2}
\left(\Delta \sigma^{j}+\bar{\Delta} \sigma^{-j} \right)
\frac{m!}{(2K)_{m}}(1+|\kappa|^2)^{-K-m}
F_{m}(|\kappa|^2:2K) \\
\qquad \qquad \qquad \qquad \qquad \mbox{where}\quad 
\kappa=\mbox{sinh}\left(C\mbox{sin}(\pi/n)\right)i,  \\
(J)\quad \frac{1}{2}
\left(\Delta \sigma^{j}+\bar{\Delta} \sigma^{-j} \right)
\frac{m!}{{}_{2J}P_m}(1-|\kappa|^2)^{J-m}
F_{m}(|\kappa|^2:2J) \\
\qquad \qquad \qquad \qquad \qquad \mbox{where}\quad 
\kappa=\mbox{sin}\left(C\mbox{sin}(\pi/n)\right)i, 
\end{array}
\right.
\end{equation}
where $C$ is a constant defined in (\ref{eq:Constants}). 

\par \noindent
Therefore let us solve (\ref{eq:sub-equation})
\begin{equation}
\label{eq:sub-equation-modified}
i\frac{d}{dt}\Psi_{0}
=\left(H_{F}+ {H_{F}}^{\dagger}\right)\Psi_{0}
=
\left\{
\left( {H_{F}}^{'}+{\tilde{H}_{F}}^{'} \right) +
\left( {H_{F}}^{''}+{\tilde{H}_{F}}^{''} \right)
\right\}\Psi_{0}
\end{equation}
with 
$
{H_{F}}^{'}+{\tilde{H}_{F}}^{'}= 
\sum_{m}\sum_{j}{\Theta}_{m,j}
\ket{\{\sigma^{j},\psi_{m}\}}\bra{\{\sigma^{j},\psi_{m}\}}
$. 
By making use of the method of constant variation again we can set 
$\Psi_{0}(t)$ as 
\begin{equation}
\label{eq:full-ansatz}
\Psi_{0}(t)=\sum_{m=0}^{\infty}\sum_{j=0}^{n-1}
\mbox{e}^{-it{\Theta}_{m,j}}a_{m,j}(t)
\ket{\{\sigma^{j},\psi_{m}\}},
\end{equation}
then we have a set of extremely complicated equations with respect to 
$\{a_{m,j}\}$. It is almost impossible to solve them. 
Therefore we consider a simple case : for $m < r$ 
\begin{equation}
\label{eq:daring-ansatz}
\Psi_{0}(t)=
\sum_{j=0}^{n-1}\mbox{e}^{-it{\Theta}_{m,j}}a_{m,j}(t)
\ket{\{\sigma^{j},\psi_{m}\}}
+ 
\sum_{j=0}^{n-1}\mbox{e}^{-it{\Theta}_{r,j}}a_{r,j}(t)
\ket{\{\sigma^{j},\psi_{r}\}}.
\end{equation}
That is, we adopt only two terms with respect to $\{m | m \geq 0\}$. 
Some careful algebras lead us to ($0\leq j \leq n-1$)
\begin{eqnarray}
\label{eq:m,r-equations}
&&i\frac{d}{dt}a_{m,j}=   \nonumber \\
&&\sum_{j^{'}=0}^{n-1}
\mbox{e}^{it\Omega (m-r)}
\mbox{e}^{it({\Theta}_{m,j}-{\Theta}_{r,j^{'}})}
\left[
\frac{\Delta}{2}
\left\{
\sum_{k=0}^{n-1}
\braa{m}
\mbox{e}^{
\frac{1}{2}
\left\{ (z_{k}-z_{k-1})L_{+}-(\bar{z}_{k}-\bar{z}_{k-1})L_{-} \right\} }
\kett{r}
\frac{\sigma^{kj-(k-1)j^{'}}}{n}
\right\}
\right.         \nonumber \\
&&\left. \qquad \qquad 
+
\frac{\bar{\Delta}}{2}
\left\{
\sum_{k=0}^{n-1}
\braa{m}
\mbox{e}^{
-\frac{1}{2}
\left\{ (z_{k}-z_{k-1})L_{+}-(\bar{z}_{k}-\bar{z}_{k-1})L_{-} \right\} }
\kett{r}
\frac{\sigma^{(k-1)j-kj^{'}}}{n}
\right\}
\right]a_{r,j^{'}}\ ,  \nonumber \\
&& \\
&&i\frac{d}{dt}a_{r,j}=   \nonumber \\
&&\sum_{j^{'}=0}^{n-1}
\mbox{e}^{it\Omega (r-m)}
\mbox{e}^{it({\Theta}_{r,j}-{\Theta}_{m,j^{'}})}
\left[
\frac{\Delta}{2}
\left\{
\sum_{k=0}^{n-1}
\braa{r}
\mbox{e}^{
\frac{1}{2}
\left\{ (z_{k}-z_{k-1})L_{+}-(\bar{z}_{k}-\bar{z}_{k-1})L_{-} \right\} }
\kett{m}
\frac{\sigma^{kj-(k-1)j^{'}}}{n}
\right\}
\right.         \nonumber \\
&&\left. \qquad \qquad 
+
\frac{\bar{\Delta}}{2}
\left\{
\sum_{k=0}^{n-1}
\braa{r}
\mbox{e}^{
-\frac{1}{2}
\left\{ (z_{k}-z_{k-1})L_{+}-(\bar{z}_{k}-\bar{z}_{k-1})L_{-} \right\} }
\kett{m}
\frac{\sigma^{(k-1)j-kj^{'}}}{n}
\right\}
\right]a_{m,j^{'}}\ . \nonumber
\end{eqnarray}
However we cannot still solve the above equations exactly, 
so let us make the {\bf rotating--wave approximation} from this stage.  
This means that the resonance condition is 
\begin{equation}
\label{eq:resonance-condition}
\Omega (m-r)+({\Theta}_{m,j}-{\Theta}_{r,j^{'}})=0
\end{equation}
for some $j$ and $j^{'}$, and we reject the remaining terms in 
(\ref{eq:m,r-equations}).  

\par \noindent 
A comment is in order. This resonance condition makes some constraint on 
the parameters $\{\omega, g, \Delta, \bar{\Delta}\}$ contained in the 
Hamiltonians (\ref{eq:Starting-Hamiltonians}), so $\omega$, $g$ and $\Delta$, 
$\bar{\Delta}$ are already not free parameters. 

\par \noindent
As for a deep meaning of this approximation see \cite{MFr5}, \cite{EFK}. 

\par \vspace{3mm} \noindent
Then we obtain relatively simple equations : 
\begin{eqnarray}
\label{eq:m,r,j,j'-equations}
&&i\frac{d}{dt}a_{m,j}= 
\left[
\frac{\Delta}{2}
\left\{
\sum_{k=0}^{n-1}
\braa{m}
\mbox{e}^{
\frac{1}{2}
\left\{ (z_{k}-z_{k-1})L_{+}-(\bar{z}_{k}-\bar{z}_{k-1})L_{-} \right\} }
\kett{r}
\frac{\sigma^{kj-(k-1)j^{'}}}{n}
\right\}
\right.         \nonumber \\
&&\left. \qquad \qquad \ 
+
\frac{\bar{\Delta}}{2}
\left\{
\sum_{k=0}^{n-1}
\braa{m}
\mbox{e}^{
-\frac{1}{2}
\left\{ (z_{k}-z_{k-1})L_{+}-(\bar{z}_{k}-\bar{z}_{k-1})L_{-} \right\} }
\kett{r}
\frac{\sigma^{(k-1)j-kj^{'}}}{n}
\right\}
\right] a_{r,j^{'}}\ ,  \nonumber \\
&&  \\
&&i\frac{d}{dt}a_{r,j^{'}}= 
\left[
\frac{\Delta}{2}
\left\{
\sum_{k=0}^{n-1}
\braa{r}
\mbox{e}^{
\frac{1}{2}
\left\{ (z_{k}-z_{k-1})L_{+}-(\bar{z}_{k}-\bar{z}_{k-1})L_{-} \right\} }
\kett{m}
\frac{\sigma^{kj^{'}-(k-1)j}}{n}
\right\}
\right.         \nonumber \\
&&\left. \qquad \qquad \ 
+
\frac{\bar{\Delta}}{2}
\left\{
\sum_{k=0}^{n-1}
\braa{r}
\mbox{e}^{
-\frac{1}{2}
\left\{ (z_{k}-z_{k-1})L_{+}-(\bar{z}_{k}-\bar{z}_{k-1})L_{-} \right\} }
\kett{m}
\frac{\sigma^{(k-1)j^{'}-kj}}{n}
\right\}
\right] a_{m,j}\ . \nonumber
\end{eqnarray}
Summing up the contents, we have 

\noindent{\bfseries General Case} ($j = j^{'}$ and $j \ne j^{'}$)\qquad 
$\Omega (m-r)+({\Theta}_{m,j}-{\Theta}_{r,j^{'}})=0$
\begin{eqnarray}
\label{eq:j,j'-equations}
&&i\frac{d}{dt}a_{m,j}= 
\frac{1}{2}\left[
{\Delta}\frac{\sigma^{j^{'}}}{n}
\left\{
\sum_{k=0}^{n-1}
\braa{m}
\mbox{e}^{
\frac{1}{2}
\left\{ (z_{k}-z_{k-1})L_{+}-(\bar{z}_{k}-\bar{z}_{k-1})L_{-} \right\} }
\kett{r}
\sigma^{k(j-j^{'})}
\right\}
\right.         \nonumber \\
&&\left. \qquad \qquad \quad
+
\bar{\Delta}\frac{\sigma^{-j}}{n}
\left\{
\sum_{k=0}^{n-1}
\braa{m}
\mbox{e}^{
-\frac{1}{2}
\left\{ (z_{k}-z_{k-1})L_{+}-(\bar{z}_{k}-\bar{z}_{k-1})L_{-} \right\} }
\kett{r}
\sigma^{k(j-j^{'})}
\right\}
\right] a_{r,j^{'}}\ ,  \nonumber \\
&&  \\
&&i\frac{d}{dt}a_{r,j^{'}}= 
\frac{1}{2}\left[
{\Delta}\frac{\sigma^{j}}{n}
\left\{
\sum_{k=0}^{n-1}
\braa{r}
\mbox{e}^{
\frac{1}{2}
\left\{ (z_{k}-z_{k-1})L_{+}-(\bar{z}_{k}-\bar{z}_{k-1})L_{-} \right\} }
\kett{m}
\sigma^{k(j^{'}-j)}
\right\}
\right.         \nonumber \\
&&\left. \qquad \qquad \quad 
+
\bar{\Delta}\frac{\sigma^{-j^{'}}}{n}
\left\{
\sum_{k=0}^{n-1}
\braa{r}
\mbox{e}^{
-\frac{1}{2}
\left\{ (z_{k}-z_{k-1})L_{+}-(\bar{z}_{k}-\bar{z}_{k-1})L_{-} \right\} }
\kett{m}
\sigma^{k(j^{'}-j)}
\right\}
\right] a_{m,j}\ . \nonumber
\end{eqnarray}

\par \noindent
Here if we define the coefficient in the equations above as  
\begin{eqnarray}
{\cal R}_{j^{'},j}=
&&{\Delta}\frac{\sigma^{j}}{n}
\left\{
\sum_{k=0}^{n-1}
\braa{r}
\mbox{e}^{
\frac{1}{2}
\left\{ (z_{k}-z_{k-1})L_{+}-(\bar{z}_{k}-\bar{z}_{k-1})L_{-} \right\} }
\kett{m}
\sigma^{k(j^{'}-j)}
\right\}  \nonumber \\
+
&&\bar{\Delta}\frac{\sigma^{-j^{'}}}{n}
\left\{
\sum_{k=0}^{n-1}
\braa{r}
\mbox{e}^{
-\frac{1}{2}
\left\{ (z_{k}-z_{k-1})L_{+}-(\bar{z}_{k}-\bar{z}_{k-1})L_{-} \right\} }
\kett{m}
\sigma^{k(j^{'}-j)}
\right\}, 
\end{eqnarray}
then 
\begin{eqnarray}
\bar{{\cal R}}_{j^{'},j}=
&&\bar{{\Delta}}\frac{\sigma^{-j}}{n}
\left\{
\sum_{k=0}^{n-1}
\braa{m}
\mbox{e}^{
-\frac{1}{2}
\left\{ (z_{k}-z_{k-1})L_{+}-(\bar{z}_{k}-\bar{z}_{k-1})L_{-} \right\} }
\kett{r}
\sigma^{k(j-j^{'})}
\right\}  \nonumber \\
+
&&{\Delta}\frac{\sigma^{j^{'}}}{n}
\left\{
\sum_{k=0}^{n-1}
\braa{m}
\mbox{e}^{
\frac{1}{2}
\left\{ (z_{k}-z_{k-1})L_{+}-(\bar{z}_{k}-\bar{z}_{k-1})L_{-} \right\} }
\kett{r}
\sigma^{k(j-j^{'})}
\right\}. 
\end{eqnarray}
To determine ${\cal R}_{j^{'},j}$ let us calculate a term 
$
\braa{r}
\mbox{e}^{
-\frac{1}{2}
\left\{ (z_{k}-z_{k-1})L_{+}-(\bar{z}_{k}-\bar{z}_{k-1})L_{-} \right\} }
\kett{m}
$ : 
\begin{eqnarray}
&&
\braa{r}
\mbox{e}^{
\frac{1}{2}
\left\{ (z_{k}-z_{k-1})L_{+}-(\bar{z}_{k}-\bar{z}_{k-1})L_{-} \right\} }
\kett{m} \nonumber \\
&&{} \nonumber \\
&&=
\left\{
\begin{array}{ll}
(N)\quad 
\sqrt{\frac{m!}{r!}}\mbox{exp}\left(-\frac{1}{2}|\kappa|^2 \right)
{L_m}^{(r-m)}\left(|\kappa|^2 \right)\kappa^{r-m} 
\sigma^{-(1/2)(r-m)}\sigma^{k(r-m)}     \\
\qquad \qquad \qquad \qquad \qquad \mbox{where}\quad 
\kappa=C\mbox{sin}(\pi/n)i,  \\
(K)\quad 
\sqrt{\frac{r!m!}{(2K)_{r}(2K)_{m}}}(1+|\kappa|^2)^{-K-(r+m)/2}
{F_{m}}^{(r-m)}(|\kappa|^2:2K)\kappa^{r-m} 
\sigma^{-(1/2)(r-m)}\sigma^{k(r-m)}     \\
\qquad \qquad \qquad \qquad \qquad \mbox{where}\quad 
\kappa=\mbox{sinh}\left(C\mbox{sin}(\pi/n)\right)i,  \\
(J)\quad 
\sqrt{\frac{r!m!}{{}_{2J}P_r {}_{2J}P_m}}(1-|\kappa|^2)^{J-(r+m)/2}
{F_{m}}^{(r-m)}(|\kappa|^2:2J)\kappa^{r-m}  
\sigma^{-(1/2)(r-m)}\sigma^{k(r-m)}     \\
\qquad \qquad \qquad \qquad \qquad \mbox{where}\quad 
\kappa=\mbox{sin}\left(C\mbox{sin}(\pi/n)\right)i, 
\end{array}
\right.
\end{eqnarray}
where $C$ is a constant defined in (\ref{eq:Constants}). 
8
\par \noindent
From this we obtain respectively 
\begin{eqnarray}
\label{eq:Rabi-frequency-N}
{\cal R}_{j^{'},j}=
&&\left\{ \Delta\frac{\sigma^{j}}{n}+
\bar{\Delta}\frac{\sigma^{-j^{'}}}{n}(-1)^{r-m} \right\}
\sqrt{\frac{m!}{r!}}\mbox{exp}\left(-\frac{1}{2}|\kappa|^2 \right)
{L_m}^{(r-m)}\left(|\kappa|^2 \right) \times
\nonumber \\
&&\kappa^{r-m}\sigma^{-(1/2)(r-m)} \sum_{k=0}^{n-1}\sigma^{k(r-m+j^{'}-j)}, \\
&&{} \nonumber \\
\label{eq:Rabi-frequency-K}
{\cal R}_{j^{'},j}=
&&\left\{ \Delta\frac{\sigma^{j}}{n}+
\bar{\Delta}\frac{\sigma^{-j^{'}}}{n}(-1)^{r-m} \right\}
\sqrt{\frac{r!m!}{(2K)_{r}(2K)_{m}}}(1+|\kappa|^2)^{-K-(r+m)/2}
{F_{m}}^{(r-m)}(|\kappa|^2:2K) \times
\nonumber \\
&&\kappa^{r-m}\sigma^{-(1/2)(r-m)} \sum_{k=0}^{n-1}\sigma^{k(r-m+j^{'}-j)}, \\
&&{} \nonumber \\
\label{eq:Rabi-frequency-J}
{\cal R}_{j^{'},j}=
&&\left\{ \Delta\frac{\sigma^{j}}{n}+
\bar{\Delta}\frac{\sigma^{-j^{'}}}{n}(-1)^{r-m} \right\}
\sqrt{\frac{r!m!}{{}_{2J}P_r {}_{2J}P_m}}(1-|\kappa|^2)^{J-(r+m)/2}
{F_{m}}^{(r-m)}(|\kappa|^2:2J) \times 
\nonumber \\
&&\kappa^{r-m}\sigma^{-(1/2)(r-m)} \sum_{k=0}^{n-1}\sigma^{k(r-m+j^{'}-j)}.  
\end{eqnarray}
Here it is easy to see 
\[
\sum_{k=0}^{n-1}\sigma^{k(r-m+j^{'}-j)}
=
\left\{
\begin{array}{ll}
0 \qquad r-m+j^{'}-j \ne 0 \ (\mbox{mod}\ n),  \\
n \qquad r-m+j^{'}-j = 0 \ (\mbox{mod}\ n), 
\end{array}
\right.
\]
so we have only to consider the case 
\quad $r-m+j^{'}-j = 0\ (\mbox{mod}\ n)$\quad 
in the following. 

Now we are in a position to solve (\ref{eq:j,j'-equations}). For simplicity 
we set ${\cal R}={\cal R}_{j^{'},j}$, then 
\begin{equation}
\label{eq:reduced equation}
i\frac{d}{dt}
\left(
\begin{array}{c}
a_{m,j} \\
a_{r,j^{'}}
\end{array}
\right)
=
\left(
\begin{array}{cc}
0& \frac{1}{2}\bar{\cal R} \\
\frac{1}{2}{\cal R}& 0
\end{array}
\right)
\left(
\begin{array}{c}
a_{m,j} \\
a_{r,j^{'}}
\end{array}
\right),
\end{equation}
so their solutions are given by 
\begin{equation}
\label{eq:reduced solution}
\left(
\begin{array}{c}
a_{m,j}(t) \\
a_{r,j^{'}}(t)
\end{array}
\right)
=
\left(
\begin{array}{cc}
\mbox{cos}(\frac{|{\cal R}|}{2}t)& 
-i\frac{\bar{\cal R}}{|{\cal R}|}\mbox{sin}(\frac{|{\cal R}|}{2}t)
\\
-i\frac{\cal R}{|{\cal R}|}\mbox{sin}(\frac{|{\cal R}|}{2}t)& 
\mbox{cos}(\frac{|{\cal R}|}{2}t)
\end{array}
\right)
\left(
\begin{array}{c}
a_{m,j}(0) \\
a_{n,j^{'}}(0)
\end{array}
\right).
\end{equation}

\par \noindent
That is, ${\cal R}$'s are just the (complex) Rabi frequencies. 
As shown in (\ref{eq:Rabi-frequency-N}), (\ref{eq:Rabi-frequency-K}), 
(\ref{eq:Rabi-frequency-J}) they are very complicated. 

\par \noindent
A comment is in order.\ If we write ${\cal R}=z=\mbox{e}^{i\theta}r$, then 
the above matrix becomes 
\begin{equation}
\left(
\begin{array}{cc}
\mbox{cos}(\frac{r}{2}t)& -i\mbox{e}^{-i\theta}\mbox{sin}(\frac{r}{2}t) \\
-i\mbox{e}^{i\theta}\mbox{sin}(\frac{r}{2}t)& \mbox{cos}(\frac{r}{2}t)
\end{array}
\right).
\end{equation}
This is a form required in Quantum Computation. However in the 2--level case 
${\cal R}$ is real, so the matrix becomes 
\[
\left(
\begin{array}{cc}
\mbox{cos}(\frac{r}{2}t)& -i\mbox{sin}(\frac{r}{2}t) \\
-i\mbox{sin}(\frac{r}{2}t)& \mbox{cos}(\frac{r}{2}t)
\end{array}
\right), 
\]
see \cite{KF15}. This is too special to perform Quantum Computation.\ 
Our result is a bonus due to the generalization from 2--level system to 
multi level one. 

\par \vspace{2mm} \noindent
We have verified the Rabi oscillations in our extended models under the 
rotating--wave approximation, so there are many things to be performed 
from Quantum Optics or Mathematical Physics. 
In the forthcoming papers we will report them.

\par \vspace{3mm}
Here let us present an (good ?) exercise to the readers : 

\noindent{\bfseries Exercise}\quad For $m < r < s$ we set 
\begin{eqnarray}
\label{eq:three-ansatz}
&&\Psi_{0}(t) \nonumber \\
=&&\sum_{j=0}^{n-1}\mbox{e}^{-it{\Theta}_{m,j}}a_{m,j}(t)
\ket{\{\sigma^{j},\psi_{m}\}}
+ 
\sum_{j=0}^{n-1}\mbox{e}^{-it{\Theta}_{r,j}}a_{r,j}(t)
\ket{\{\sigma^{j},\psi_{r}\}}
+ 
\sum_{j=0}^{n-1}\mbox{e}^{-it{\Theta}_{s,j}}a_{s,j}(t)
\ket{\{\sigma^{j},\psi_{s}\}}.   \nonumber 
\end{eqnarray}
Solve the equation (\ref{eq:sub-equation-modified}) under this ansatz. 
Then you must pay attention to the consistency of some resonance conditions 
like (\ref{eq:resonance-condition}).

\section{Unitary Operations in Quantum Computation}

In this section we make a brief comment on application of the results 
in the preceding section to quantum computation based on qudits space. 
Remind once more that the following arguments are based on the {\bf 
rotating wave approximation}. 

First we must solve the equation which $j$ and $j^{'}$ satisfy 
\begin{equation}
j^{'}-j+r-m=0\ (\mbox{mod}\ n)\quad \mbox{for}\quad m < r. 
\end{equation}
The solutions are easily given as follows. 
\begin{eqnarray}
(j^{'},\ j)&=&(n-r+m,\ 0),\ (n-r+m+1,\ 1),\ \cdots,\ (n-1,\ r-m-1), 
\nonumber \\ 
           & &(n-r+m-1,\ n-1),\ (n-r+m-2,\ n-2),\ \cdots,\ (0,\ r-m).
\end{eqnarray}

\par \noindent 
Here let us rewrite the equation (\ref{eq:reduced equation}) in a full form 
\begin{equation}
i\frac{d}{dt}
\left(
\begin{array}{c}
a_{m,0} \\
\cdot \\
a_{m,j} \\
\cdot \\
a_{m,n-1} \\
a_{r,0} \\
\cdot \\
a_{r,j^{'}} \\
\cdot \\
a_{r,n-1} 
\end{array}
\right)=
\left(
\begin{array}{cccccccccc}
 0& & & & & & & & & \\
 & \cdot& & & & & & & & \\
 & & 0& & & & & \frac{1}{2}\bar{\cal R}& & \\
 & & & \cdot& & & & & & \\
 & & & & \cdot& & & & & \\
 & & & & & \cdot& & & & \\
 & & & & & & \cdot& & & \\
 & & \frac{1}{2}{\cal R}& & & & & 0& & \\
 & & & & & & & & \cdot& \\
 & & & & & & & & & 0 
\end{array}
\right)
\left(
\begin{array}{c}
a_{m,0} \\
\cdot \\
a_{m,j} \\
\cdot \\
a_{m,n-1} \\
a_{r,0} \\
\cdot \\
a_{r,j^{'}} \\
\cdot \\
a_{r,n-1} 
\end{array}
\right)
\end{equation}
where ${\cal R}={\cal R}_{j^{'},j}$. 
Then the solution (corresponding to (\ref{eq:reduced solution})) is explicitly 
\begin{eqnarray}
\label{eq:unitary-transformation}
&&\left(
\begin{array}{c}
a_{m,0}(t) \\
\cdot \\
a_{m,j}(t) \\
\cdot \\
a_{m,n-1}(t) \\
a_{r,0}(t) \\
\cdot \\
a_{r,j^{'}}(t) \\
\cdot \\
a_{r,n-1}(t) 
\end{array}
\right)  \nonumber \\
=&&
\left(
\begin{array}{cccccccccc}
 1& & & & & & & & & \\
 & \cdot& & & & & & & & \\
 & & \mbox{cos}(\frac{|{\cal R}|}{2}t)& & & & & 
 -i\frac{\bar{\cal R}}{|{\cal R}|}\mbox{sin}(\frac{|{\cal R}|}{2}t)& & \\
 & & & \cdot& & & & & & \\
 & & & & \cdot& & & & & \\
 & & & & & \cdot& & & & \\
 & & & & & & \cdot& & & \\
 & & -i\frac{{\cal R}}{|{\cal R}|}\mbox{sin}(\frac{|{\cal R}|}{2}t)& & & 
 & & \mbox{cos}(\frac{|{\cal R}|}{2}t)& & \\
 & & & & & & & & \cdot& \\
 & & & & & & & & &1  
\end{array}
\right)
\left(
\begin{array}{c}
a_{m,0}(0) \\
\cdot \\
a_{m,j}(0) \\
\cdot \\
a_{m,n-1}(0) \\
a_{r,0}(0) \\
\cdot \\
a_{r,j^{'}}(0) \\
\cdot \\
a_{r,n-1}(0) 
\end{array}
\right). 
\end{eqnarray}

\par \noindent 
That is, we have a unitary operation which will be written as $U(j,j^{'};t)$. 
This operation is crucial in Quantum Computation. 

\par \noindent
As a result we obtained $n$ elementary unitary operations 
\begin{eqnarray}
\label{eq:n-unitary operattions}
&&U(0,n-r+m;t),\ U(1,n-r+m+1;t),\ \cdots,\ U(r-m-1,n-1;t), \nonumber \\
&&U(r-m,0;t),\ U(r-m+1,1;t),\ \cdots,\ U(n-1,n-r+m-1;t). 
\end{eqnarray}
One can combine these operations suitably to obtain the (complex) operation 
required in Quantum Computation.

We would like to apply these operations to Quantum Computation on the space 
of 2--qudits. For that we need some modification in the preceeding section. 
Namely, we must consider a set of $n$ states : 
for $m_{0} < m_{1} < \cdots < m_{n-1}$
\begin{equation}
\label{eq:n-pairs ansatz}
\Psi_{0}(t)=\sum_{k=0}^{n-1}\sum_{j=0}^{n-1}
\mbox{e}^{-it{\Theta}_{m(k),j}}a_{m(k),j}(t)\ket{\{\sigma^{j},\psi_{m(k)}\}}
\end{equation}
instead of (\ref{eq:daring-ansatz}). Of course for any pair $m_{k}<m_{l}$  
we have the same result. We cannot solve the equation at a stroke, so we 
reduce the calculation on $n$--system to that of the pairs. 
Remind once more that in this paper we are interested in Rabi oscillations 
between two states. By combining these Rabi oscillations one after another 
we obtain unitary operations in our full system. 

\par \noindent 
For (\ref{eq:n-pairs ansatz}) we have the huge (block-) equation 
\begin{equation}
\left(
\begin{array}{c}
{\bf a_{m(0),j}}(t) \\
{\bf a_{m(1),j}}(t) \\
\cdot \\
\cdot \\
\cdot \\
{\bf a_{m(n-1),j}}(t) 
\end{array}
\right)= 
\left(
\begin{array}{ccccccc}
A_{00}& A_{01}& \cdot& \cdot& \cdot& A_{0,n-1} \\
A_{10}& A_{11}& \cdot& \cdot& \cdot& A_{1,n-1} \\
\cdot & \cdot &      &      &      & \cdot     \\
\cdot & \cdot &      &      &      & \cdot     \\
\cdot & \cdot &      &      &      & \cdot     \\
A_{n-1,0}& A_{n-1,1}& \cdot& \cdot& \cdot& A_{n-1,n-1} \\
\end{array}
\right)
\left(
\begin{array}{c}
{\bf a_{m(0),j}}(0) \\
{\bf a_{m(1),j}}(0) \\
\cdot \\
\cdot \\
\cdot \\
{\bf a_{m(n-1),j}}(0) 
\end{array}
\right) 
\end{equation}
where each ${\bf a_{m(k),j}}$ is a $n$ vector and 
$A_{ij}$ a $n\times\ n$ matrix. 

\par \noindent 
For example, if we pay attention to the pair $m_{k}<m_{l}$ then the block 
(in the above huge matrix) 
\begin{equation}
\left(
\begin{array}{cc}
A_{m(k)m(k)}& A_{m(k)m(l)} \\
A_{m(l)m(k)}& A_{m(l)m(l)} 
\end{array}
\right)
\end{equation}
is just one of above unitary operations in (\ref{eq:n-unitary operattions}) 
($m=m_{k},\ r=m_{l}$) and the remainders are $A_{pp}={\bf 1}$, 
$A_{pq}={\bf 0}\ (p\ne q)$. 
As a total we have the elementary unitary transformations in $U(n^{2})$ 
with the following number 
\[
n{}_nC_2=\frac{n^{2}(n-1)}{2}. 
\]

\par \noindent 
Here if we identify (\ref{eq:n-pairs ansatz}) with an element on the space 
of 2--qudits like 
\begin{equation}
\sum_{k=0}^{n-1}\sum_{j=0}^{n-1}
a_{k,j}(t)\ket{k}\otimes \ket{j}\ \in \ \fukuso^{n}\otimes \fukuso^{n} 
\end{equation}
where $\fukuso^{n}=\mbox{Vect}_{\fukuso}\{\ket{0},\ket{1},\cdots,\ket{n-1}\}$, 
then we have plenty of unitary transformations in $U(n^{2})$. We want to 
construct or perform some logic gates on the space of 2--qudits, 
see \cite{KF16}, \cite{KuF} for details.

\par \noindent 
For example, we would like to construct the (reverse) controlled--shift gate 
((reverse) controlled--unitary ones more generally) by combining 
elementary unitary operations obtained in this paper : 
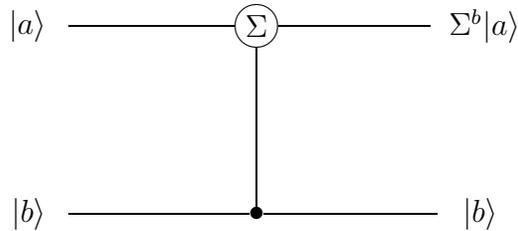
\begin{center}
\setlength{\unitlength}{1mm}  
\begin{picture}(120,50)
\put(25,35){\line(1,0){22}}   
\put(13,30){\makebox(9,10)[r]{$|a\rangle$}} 
\put(53,35){\line(1,0){20}}   
\put(76,30){\makebox(9,10)[r]{$\Sigma^{b}|a\rangle$}} 
\put(25,10){\line(1,0){24}}   
\put(13, 5){\makebox(9,10)[r]{$|b\rangle$}} 
\put(51,10){\line(1,0){23}}   
\put(73, 5){\makebox(9,10)[r]{$|b\rangle$}} 
\put(50,11){\line(0,1){21}}   
\put(47,5){\makebox(6,10){$\bullet$}} 
\put(50,35){\circle{6}}               
\put(47,30){\makebox(6,10){$\Sigma$}}         
\end{picture}
\end{center}
where $\Sigma$ in the above figure is just $\Sigma_{1}$ in (\ref{Sigma-1}), 
and $a$ and $b$ run from $0$ to $n-1$.

\par \vspace{3mm} 
We would like to conclude this section by stating that a possible realization 
of our model could be found in Josephson junctions \cite{NPT} and 
ion cavities \cite{several-1}, \cite{several-2}, \cite{several-3}.

\section{Discussion}

In this paper we extended the works in \cite{MFr}, \cite{MFr3} and \cite{KF15} 
based on 2--level system to n--level system and obtained the general theory of 
Rabi frequencies in the strong coupling regime under the rotating--wave 
approximation. The essence of our work is that Rabi frequencies are in this 
case deeply connected to the matrix elements of coherent and/or generalized 
coherent operators based on $su(2)$ and $su(1,1)$, \cite{KF14}, \cite{KF6}. 

One of motivations of this study is to apply our results to 
Holonomic Quantum Computation developed by Italian group (Pachos, 
Rasetti and Zanardi) and the author, see \cite{ZR}, \cite{PZR}, \cite{PC}, 
\cite{PZ} and \cite{KF0}---\cite{KF4}. 

Holonomic Quantum Computation is based on 2--level system at this stage, but 
it is very natural to extend it from 2--level system to n--level system 
(let us call it Generalized Holonomic Quantum Computation (GHQC)). 
In fact the author has attempted its extension in \cite{KF0}, but it was 
not complete due to some lack of new method.

As shown in this paper we have obtained some new method or technique, so 
we would like to challenge once more.

\par \vspace{10mm}
\noindent{\em Acknowledgment.}\\
The author wishes to thank Marco Frasca and Zakaria Giunashvili 
for their helpful comments and suggestions.


\end{document}